\documentclass[10pt,letterpaper]{article}
\usepackage{opex3}
\usepackage{amsmath,amssymb,amsfonts} 
\usepackage{quotmark}
\usepackage{xspace}
\usepackage{citesort}
\usepackage{color}

\newcommand{\ulm}{u_{lm}}
\newcommand{\vlm}{u_\text{\tiny 2}}
\newcommand{\wlm}{w_\text{\tiny 3}}

\newcommand{\uii}{u_\text{\tiny 11}}
\newcommand{\ui}{u_\text{\tiny 1}}
\newcommand{\eibzt}{e^{i(\beta z - \omega t)}}
\newcommand{\Jlmi}{J_{l-1}(\ui r)}
\newcommand{\Jlpi}{J_{l+1}(\ui r)}
\newcommand{\Jl}{J_l(\ui r)}
\newcommand{\Jlmlm}{J_{l-1}(\ulm r)}
\newcommand{\Jlplm}{J_{l+1}(\ulm r)}

\newcommand{\sinlp}{\sin(l \phi + \varphi) }
\newcommand{\coslp}{\cos(l \phi + \varphi) }
\newcommand{\eilp}{\exp(i l \phi)}
\newcommand{\Er}{E_r}
\newcommand{\Ez}{E_z}
\newcommand{\Ephi}{E_\phi}
\newcommand{\Hr}{H_r}
\newcommand{\Hz}{H_z}
\newcommand{\Hphi}{H_\phi}
\newcommand{\cl}{\text{cl}}
\newcommand{\Ecllm}{E_{lm}}
\newcommand{\Ccllm}{C_{lm}}

\newcommand{\Plm}{P_{lm}}
\newcommand{\Ulm}{U_{lm}}
\newcommand{\vtwoone}{v_\text{\tiny 21}}
\newcommand{\vthreetwo}{v_\text{\tiny 32}}

\newcommand{\neff}{\bar{n}}
\newcommand{\ncore}{n_{\text{\tiny 1}}}
\newcommand{\nclad}{n_{\text{\tiny 2}}}
\newcommand{\nair}{n_{\text{\tiny 3}}}

\newcommand{\micron}{\textmu m\xspace}
\newcommand{\mLP}[2]{$\mbox{LP}_{#1,#2}$}
\newcommand{\mHE}[2]{$\mbox{HE}_{#1,#2}$}
\newcommand{\mEH}[2]{$\mbox{EH}_{#1,#2}$}
\newcommand{\acore}{a_{\text{\tiny 1}}}
\newcommand{\aclad}{a_{\text{\tiny 2}}}

\newcommand{\bE}{\mathbf{E}}
\newcommand{\bH}{\mathbf{H}}

\newcommand{\kappacoco}{\kappa_{11}}
\newcommand{\kappacocl}{\kappa_{lm}}
\newcommand{\Aco}{A_{11}}
\newcommand{\Bco}{B_{11}}
\newcommand{\Bcl}{B_{lm}}
\newcommand{\dd}{\mathrm{d}}
\newcommand{\dz}{\mathrm{d}z}
\newcommand{\deltacoco}{\delta_{11}}
\newcommand{\deltacocl}{\delta_{lm}}

\begin{document}

\title{Cladding mode coupling in highly localized fiber Bragg gratings:
modal properties and transmission spectra}

\author{Jens Thomas$^{1*}$, Nemanja Jovanovic$^2$, Ria G. Becker$^1$,\\ Graham D.  Marshall$^2$, Michael J.  Withford$^2$,\\ Andreas  T\"unnermann$^1$, Stefan Nolte$^1$ and M. J.  Steel$^2$}

\address{$^1$ Friedrich-Schiller University, Institute of Applied Physics,\\ Max-Wien Platz 1, 07743 Jena, Germany\\
$^2$
MQ Photonics Research Centre and  \\
Centre for Ultrahigh bandwidth Devices for Optical Systems (CUDOS),
Department of Physics and Astronomy,\\
Macquarie University, North Ryde, New South Wales 2109, Australia
}

\email{* thomas@iap.uni-jena.de} 



\begin{abstract}
The spectral characteristics of a fiber Bragg grating (FBG) with a transversely
inhomogeneous refractive index profile, differs considerably from that of a
transversely uniform one.  Transmission spectra of inhomogeneous and
asymmetric FBGs that have been inscribed with focused ultrashort pulses with
the so-called point-by-point technique are investigated.  The cladding mode
resonances of such FBGs can span a full octave in the spectrum and are very
pronounced (deeper than 20dB).  Using a coupled-mode approach, we compute the strength of
resonant coupling and find that coupling into cladding modes of higher
azimuthal order is very sensitive to the position of the modification in the
core.  Exploiting these properties allows precise control of such reflections
and may lead to many new sensing applications.  
\end{abstract} 


\bibliographystyle{osajnl}

\section{Introduction}
In a typical transmission spectrum of fiber Bragg gratings (FBGs), it is common
to observe a comb of resonances on the shorter wavelength side of the Bragg
peak.  These dips result from the resonant coupling between the forward-propagating 
core mode and modes that propagate backwards within
the cladding of the fiber~\cite{Erdogan:1997p833,Mizrahi:1993p798}.  Since
modes guided in the cladding of a fiber are very sensitive to bending of the
fiber as well as the surrounding environment, they are widely exploited for
sensing applications.  In such sensor arrangements, FBGs and long period
gratings (LPGs) are used as core to cladding mode
converters~\cite{Jauregui:2004p3427}.  However, conventional FBGs, in which the
cross-section of the core is uniformly modified, couple light
only into cladding modes of the lowest azimuthal order~\cite{Erdogan:1997p833}.
In general, coupling into higher azimuthal fiber modes can be realized by FBGs
that are tilted~\cite{Erdogan:1996p855,Chan:2007p7575,Zhou:2006p3488} or have
an asymmetric transverse index profile~\cite{Mizunami:2000p2676}.  More
elaborate in-fiber devices to achieve strong cladding mode resonances involve
tapered fiber sections~\cite{Guo:2009p7572} or deliberately misaligned fiber
splices~\cite{Guo:2008p6002}.


In this paper, we investigate how strong cladding mode coupling can be achieved and controlled with highly localized FBGs.
Such FBGs can be fabricated using focused ultrashort lasers operating in the
near infrared (NIR)~\cite{Martinez:2004p2878}.  Due to the nonlinear nature of the
absorption process of light pulses in the femtosecond regime, highly localized
refractive index profiles can be obtained in non-photosensitive materials
without damaging the surrounding fiber material. The induced refractive index
change is no longer confined to the photosensitive region of the fiber (as is
the case for FBGs written with ultra-violet lasers), but is determined solely by the focusing
geometry of the ultrashort laser.  Both the
size and position of the modifications depend heavily on the alignment and
focusing of the writing beam~\cite{Marshall:2010p8260,Thomas:2007p80}. 
Highly localized FBG exhibit a pronounced coupling to cladding modes.
Our aim is to investigate and understand these cladding mode spectra in detail in terms of a full vectorial coupled mode theory.
Our work shows that excitation of cladding-modes with strong selectivity and precision is possible,
detailing how the coupling behavior is determined by the fiber geometry and to which extend it can be manipulated 
with the cross-section of the FBG.

\section{Realization and spectra of highly localized FBG}\label{sec:exp}
\subsection{Fabrication and characterization}
The details of the grating inscription procedure closely follow that, which has been reported in~\cite{Marshall:2010p8260,Jovanovic:2009p2656}.  A femtosecond laser
(Spectra-Physics, Hurricane) that delivered pulses shorter than $110$~fs at a
wavelength of $800$~nm was utilized for grating inscription.  A $20\times$ oil
immersion objective ($NA=0.8$) was used for focusing the ultrashort pulses
into the core of the fiber, as well as for imaging the fiber core before and
after inscription of the gratings. All experiments were performed with a
standard single mode fiber (Corning SMF-28e). The fiber was completely stripped of its polymer
cladding prior to grating inscription and subsequently threaded through a glass
ferrule which was placed in close proximity to the objective lens.  The lens,
ferrule and fiber were immersed in an index matching oil in order to minimize
refraction as the pulse propagates towards the focal spot.  The ferrule was
positioned such that the core of the fiber was located at the focus of the
laser beam and the fiber was then pulled through the ferrule at a constant
velocity while the grating was inscribed.  By adjusting the ferrule's
location with respect to the focus of the lens, the transverse position of the
modifications of the FBG could be controlled to micrometer accuracy within the
fiber core.  The error in the longitudinal position of the modifications was
much smaller, because the fiber was pulled by an air-bearing translation stage
designed for constant velocity traversals, and the repetition rate of
the laser was constant. All gratings were inscribed in SMF-28e fibers and
are $20$~mm long.  Except where stated otherwise, the gratings have a period $\Lambda=1.062$~\textmu m 
that reflects 1.5~\textmu m light in the second order.

\begin{figure}[htbp]
	\centering
		\includegraphics[width=\textwidth]{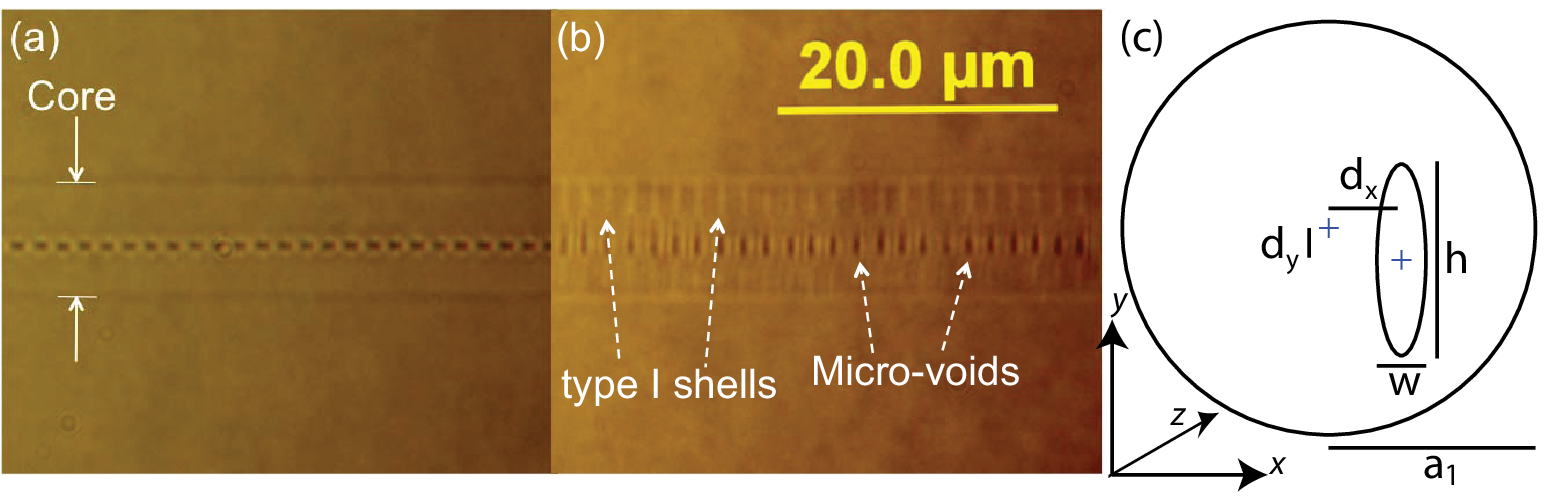}
  \caption{
Transmission differential interference contrast microscope images of the index
modifications viewed from (a) above (parallel to the
direction of the writing laser beam) and (b) the side (perpendicular to the
direction of the writing laser  beam).
(c) Schematic diagram of the cross-section of the femtosecond induced
modifications within the core of radius $\acore$, establishing the coordinate system, where
$z$ points along the fiber and $x$ and $y$ are perpendicular to the fiber axis.
} 
\label{fig:mod_sketch}
\end{figure}

Pulse energies between $200$ and $275$~nJ were used, which provided enough
irradiance within the focal volume to cause a micro-explosion in the fiber,
leaving behind a micro-void and hence forming a type II-IR Point-by-Point (PbP) FBG~\cite{Smelser:2005p41}.  
The micro-voids are surrounded by a compressed region of increased refractive
index.  In previous work, we have demonstrated that the net effective refractive index change
is negative, although this is not significant in the present
work~\cite{Jovanovic:2009p2656}.  
Microscope images of the grating modifications taken from the top (a) and side (b) views
are shown in Fig.~\ref{fig:mod_sketch}.  The micro-void regions, which are
highlighted in Fig.~\ref{fig:mod_sketch}, measure 0.4~\micron in width by 
1.9~\micron in height. They are enclosed by a densified Type I~shell of width approximately 1~\micron and height 8~\micron,
where the index is slightly increased. The modifications are not necessarily centered, thus having an
offsets of $d_x$ and $d_y$.

\subsection{Spectral response}
\begin{figure}[htbp]
\begin{center}
\includegraphics[width=\textwidth]{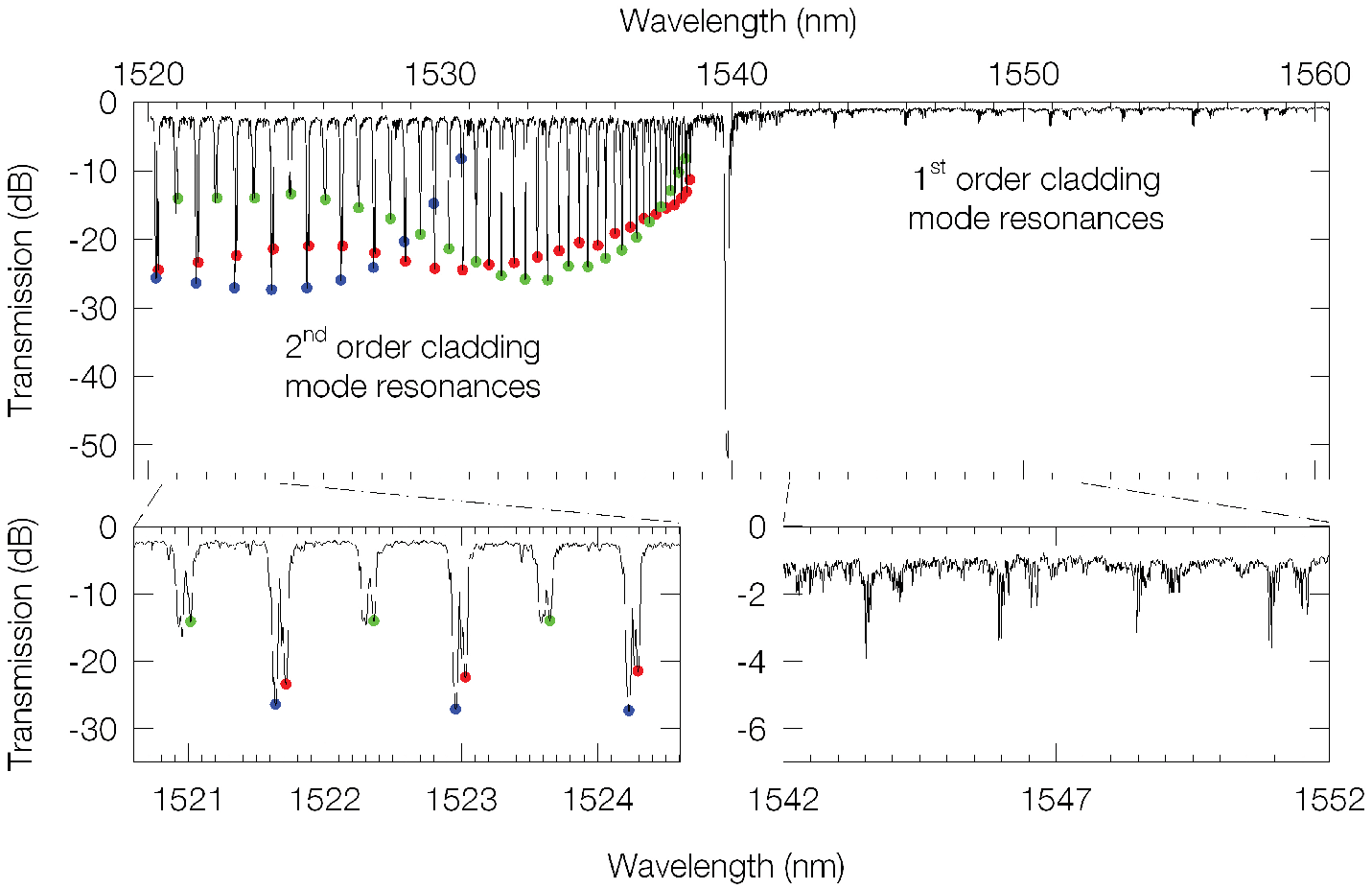}
\caption{Transmission spectrum of a PbP grating operating in second order at 1540~nm and showing strong cladding mode resonances. According to the notation laid out in section \ref{sec:classify}, the colored dots indicate the envelopes of resonances with HE $l=1$ (red), EH $l=1$ (blue) and HE $l=2$ (green).The lower graphs show expanded scales to illustrate the HE/EH splitting of modes with high radial order $m$ (lower left), and the existence of first-order cladding mode peaks right up to the second-order Bragg peak (lower right). 
}
\label{fig:firstpeaks} 
\end{center} 
\end{figure}

The measurement of the transmission and reflection properties was carried out with
a swept wavelength system (SWS) (JDS Uniphase), that consisted of a narrow
line-width external cavity laser diode.  The probing wavelength of
the laser diode can be continuously swept from $\lambda=1520$~to~$1570$~nm with a
resolution of $3$ pm.  
A typical transmission spectrum is shown in Fig.~\ref{fig:firstpeaks}.
In contrast to FBGs written with conventional
techniques, the cladding mode structure is very pronounced: the reflectivity of
core light into cladding modes is higher than $70$~percent, even at wavelengths
$20$~nm shorter than the Bragg wavelength $\lambda_B=1540$~nm. For gratings at 1555~nm, we see strong resonances 35~nm below the Bragg peak, limited only by the wavelength
range of our measurement system. Comparable cladding mode
spectra can be obtained with conventional techniques only with blazed FBGs with
large tilt angles, e.g.  80\textdegree~\cite{Zhou:2006p3488}, whereas our gratings are untilted.  

As detailed in section \ref{sec:classify}, each resonance can be attributed to the coupling of the core
mode to one or more definite cladding modes characterized by their radial and azimuthal mode numbers.  
It can be seen that the spectrum contains three sets of resonances with distinct well-defined envelopes.
We will show that the envelopes correspond to modes with even (green dots) or odd (red dots) azimuthal mode numbers.


%

Close inspection reveals that many of the resonances, 
especially at wavelengths lower than 1530 nm (see lower left panel of Fig.~\ref{fig:firstpeaks}), 
actually consist of doublets (marked by red and blue dots
respectively).  No such splitting is observed at the low order (long wavelength) resonances. 
These doublets result from the hybrid nature of the cladding guided modes, which will be discussed in detail
 in section \ref{sec:classify}.

Finally, the lower right panel of Fig.~\ref{fig:firstpeaks} provides a striking illustration of the strength
of the cladding mode coupling in these gratings.
It shows the response at wavelengths longer than the second-order Bragg peak at 1540~nm.
A sequence of regularly-spaced pairs of resonances 1--2~dB deep, each themselves
with a recurring splitting pattern, is clearly
visible.  We interpret these peaks as the tail of the first-order cladding mode
spectrum, and they therefore correspond to resonances many hundreds of modes
below the fundamental.  The actual first order Bragg peak would occur
in the mid-IR around 3~\micron and is beyond our observable range due to losses
and the spectral range of our near-IR characterization setup.

In the following sections, we perform calculations to understand these types of
cladding mode spectra and explain their features in precise detail. 

\section{Classification and properties of modes in three layer fibers} \label{sec:classify}
For the purpose of calculating and classifying the cladding modes we
consider a standard, cylindrical step-index fiber supporting a single
core-guided mode over the wavelength range of interest.
The elliptical modifications introduce a small anisotropy
to the core and the modes are technically birefringent, but this effect
is insignificant for the cladding modes.
We therefore treat the fiber as having perfect circular symmetry, 
with core radius $\acore$ and cladding radius $\aclad$.  
The refractive indices are $\ncore, \nclad$, respectively.  

Exploiting the small difference in $\ncore$ and $\nclad$,
cladding modes are most often calculated with a two-layer model
consisting of a solid rod of index $\nclad$ and radius $\aclad$
surrounded by an air cladding with index 1.  They are further often 
treated in the approximation of weak-guidance or linearly polarized (LP) 
modes~\cite{Snyder:1978p2682}.
This approximation is usually
sufficient for computing the position of the resonances in the spectrum
\cite{Hewlett:1995p2660,Hewlett:1996p2659}, since the propagation constants of
the true HE and EH hybrid modes of the step index fiber 
differ only by a very small amount $\Delta \beta$~\cite{Snyder:1978p2682}.

It has previously been found, however, that in order to accurately calculate coupling coefficients 
between core and cladding modes, both these approximations must be avoided~\cite{Erdogan:1997p833}.  
The core-cladding boundary influences the 
electric field distribution of the lower order modes especially.
Hence it is necessary to solve the cladding modes for the full three-layer
structure~\cite{Tsao:1989p1975}.  
Moreover, because fiber modes of higher azimuthal order are not linearly polarized, 
we need a full vectorial approach~\cite{Snyder:1978p2682}
to accurately explain splittings in the cladding-mode spectra that do not appear in the LP picture.



In the remainder of the paper, the core
radius $a_{1}=4.15$~\micron and the cladding radius $a_{2}=$ 62.5~\micron.  The
refractive indices of the core, cladding and surrounding air
are $\ncore=1.4670, \nclad=1.4618$ and $n_3=1.0$ respectively.  
Modes travel in the $z$ direction with the propagation constant $\beta=2\pi
\neff/ \lambda$, where $\neff$ is the effective index.  
A mode is guided within the core if $\neff>\nclad$, and in the cladding if $\nclad>\neff>n_3$.
To excite the system, we launch a single linearly polarized \mHE{1}{1} core mode.

\subsection{Field expressions for the modes}
In our analysis, we make  extensive use of the three-layer fiber model in the form laid out by
Erdogan~\cite{Erdogan:1997p833}.  He adapted the full
analytical solution of the three-layer treatment of
Tsao~\cite{Tsao:1989p1975}, which yields all hybrid  modes with the azimuthal
and radial integer indices $l$ and $m$.  However, in \cite{Erdogan:1997p833}, Erdogan
focused solely on cladding mode reflections of conventional fiber gratings,
for which only $l=1$ resonances occur.  In our work, higher order
modes with $l>1$ are needed.  
We use the notation employed by Erdogan, but express the azimuthal
dependence in trigonometric rather than exponential form.
In cylindrical coordinates ($r$, $\phi$, $z$), the electric $\bE$ and magnetic $\bH$ fields of the cladding modes inside the core
($r<\acore$) can be expressed in terms of Bessel functions $J_n$ of the first kind
\begin{subequations}
\label{eq:tsaocore}
\begin{alignat}{2}
\Ez    &=&    &\Ecllm \frac{\ui^2}{\beta} P \Jl \sinlp \eibzt \\
\Er    &=& i &\Ecllm \frac{\ui}{2} \left[ (1-P) \Jlmi + (1+P ) \Jlpi \right] \sinlp \eibzt \\
\Ephi  &=& i &\Ecllm \frac{\ui}{2} \left[ (1-P) \Jlmi - (1+P ) \Jlpi \right] \coslp \eibzt \\
\Hz    &=&    &\Ecllm \frac{\neff}{Z_0}\frac{\ui^2}{\beta} \Jl \coslp \, \eibzt \\
\Hr    &=& i &\Ecllm \frac{\neff}{Z_0} \frac{\ui}{2} \left[ - \left(1-P\frac{\ncore^2}{\neff^2}\right) \Jlmi +\left(1+P\frac{\ncore^2}{\neff^2} \right) \Jlpi \right] \coslp \, \eibzt  \\
\Hphi  &=&  -i &\Ecllm \frac{\neff}{Z_0} \frac{\ui}{2} \left[ - \left(1-P\frac{\ncore^2}{\neff^2}\right) \Jlmi -\left(1+P\frac{\ncore^2}{\neff^2} \right) \Jlpi \right] \sinlp \, \eibzt ,
\end{alignat}
\end{subequations}
with the transverse wavevector $\ui=(2\pi/\lambda) \sqrt{\ncore ^2- \neff^2}$. The constant 
$Z_0=\sqrt{\mu_0/\epsilon_0}\approx 376.7\Omega$ is the electromagnetic impedance in vacuum.
Note that in these expressions $\neff, \beta, \ui$ and $P$ all depend on the mode
indices $l$ and $m$, and we sometimes indicate this explicitly by including
the mode indices as subscripts.  In particular, the mode parameter
\begin{equation}\label{P}
P=P_{lm}=-\frac{\neff_{lm}  i\zeta_0}{n_{1}^2}
\end{equation}
characterizes the relative strength of the longitudinal field components,
and is used to classify modes as HE or EH.  The imaginary
parameter $\zeta_0$ is determined by solving the dispersion equation for the three-layer fiber.
The normalization constant $E_{lm}^{cl}$ is set such that all modes carry the same power.  
Complete expressions for the dispersion relation and the cladding mode fields in the regions $r>\acore$
are provided in appendices A and B respectively.  

Note that Eqs.~\eqref{eq:tsaocore} represent two orthogonal sets of solutions, 
distinguished by the rotation angle $\varphi$, which we take as $\varphi=0$ or $\varphi=-\pi /2$. 
Thus, all hybrid mode solutions of the form in Eqs.~\eqref{eq:tsaocore} appear
as degenerate pairs of fields with orthogonal polarization states. Since in our experiments
the input mode \mHE{1}{1} is $y$-polarized, it is represented purely by the $\varphi=0$ state.
In general, a grating may produce coupling to cladding modes
of both $\varphi=0$ and $\varphi=-\pi /2$.  As discussed below however, careful alignment
of the polarization as well as the defect orientation and location eliminates coupling to the 
perpendicular $\varphi=-\pi/2$ modes.

\subsection{Classification of the modes}\label{sec:classification}
It is useful to recall some standard properties of the core-guided hybrid modes of the two-layer model.
In that system, the modes are classified as HE or EH according to the value of $P_{lm}$.  
The HE modes have $-1<P_{lm}<0$ and the EH modes have $P_{lm}>1$,
with $P_{lm} \approx \pm1 $ for the strongly bound modes,
(and $P_{lm}=\pm 1 $ in the LP limit).
For azimuthal number $l$, the transverse field components of the HE modes 
have predominantly a $J_{l-1}(\ui r)$ radial dependence,
whereas the EH modes are dominated by the $J_{l+1}(\ui r)$ Bessel function.
Consequently, a mode \mHE{l}{m} has a time-averaged Poynting flux $\langle S_z \rangle$
with peaks closer to the origin than the corresponding \mEH{l}{m} mode.
For fixed $l$, the modes occur in a strict alternating sequence of HE and EH,
(excepting the fundamental \mHE{1}{1}.

The mode parameter $\Plm$ also turns 
out to be very convenient for the classification of cladding modes in the
three-layer fiber.
However, the HE/EH criterion has to be modified.
For further discussion, we define the projected normalized wavevector
\begin{equation}
U_{lm} = \frac{2\pi \acore}{\lambda} \frac{\ncore}{\nclad} \sqrt{\nclad^2-\neff_{lm}^2},
\end{equation}
which corresponds to the normalized frequency a cladding mode would
have in the core region if the refraction on the core-cladding boundary is
ignored.  

\begin{figure}[htbp]
	\centerline{ \includegraphics[width=0.9\textwidth]{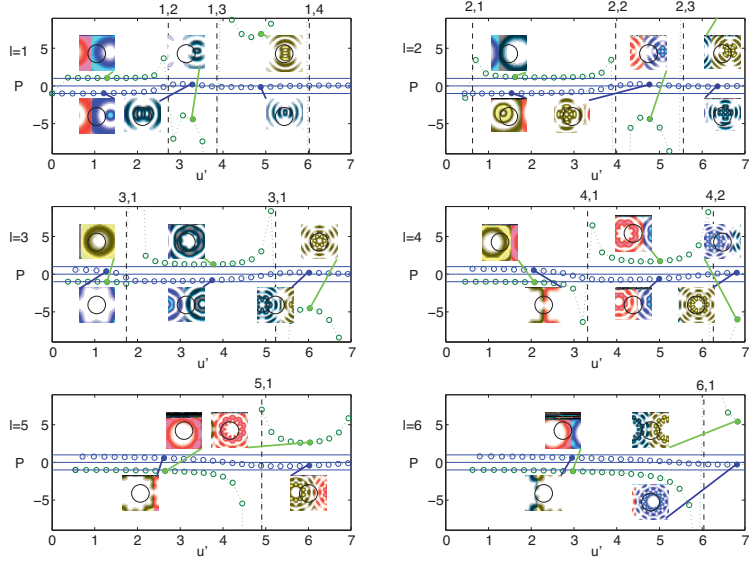} }
	\caption{Hybrid mode parameter $P_{lm}$ for $l=1$ to $l=6$ as a function of  the 
projected normalized wavevector $U_{lm}$.  Blue and green circles denote modes labeled HE and EH respectively. 
The insets display the intensity of the cladding modes in the region $r<3a_1$ with the black circle denoting
the core boundary (EH modes are shown above the $x$-axis, HE modes below it).  
Vertical lines indicate the virtual cutoffs $\tilde{U}_{lm'}$ and are labelled on the upper edge of each
graph.  The change in character of the modes at each virtual cutoffs is clear.}
	\label{fig:p_multimode}
\end{figure}

In Fig.~\ref{fig:p_multimode}, we plot the hybrid mode parameter $P_{lm}$
against the normalized wavevector for the lowest six azimuthal
orders.  The plots show that $P_{lm}$ has a complex dependence and
is not restricted to the ranges $-1<P<0$ and $P>1$.
Adjusting the two-layer fiber definitions,
we designate the modes with $|\Plm|<1$ as HE (colored blue) and those with  $|\Plm|>1$ 
as EH (colored green).  
Note that with this definition, we retain consistency with the classification of 
core modes in the two-layer picture, and the modes again appear in a strict
sequence alternating between HE and EH.
Now in Fig.~\ref{fig:p_multimode} we find that for each value of $l$, 
the diagrams can be divided into regions according to
the sign of $P_{lm}$ for either set of modes.  In fact, the transitions,
indicated by vertical dashed lines, correspond to mode indices $m$ at which an additional
intensity ring enters the core.  
We refer to these discontinuities as ``virtual
cut-offs'' $\tilde{U}_{lm'}$, where the virtual radial number $m'$ is to be
associated with the number of maxima within the core. The cutoff value
$\tilde{U}_{lm'}$ is the value of $U_{lm}$ at which the $m'$th maximum appears in
the core.
The intensity distributions near and in the fiber core
$S_z=\frac{1}{2}(E_rH_{\phi}^+-E_{\phi}H_r^*)$ for a number of modes are plotted as insets. 
It can indeed be seen that a further intensity ring appears within the 
core each time there is a discontinuity in the $P_{lm}$ plot. 

The character of the modes changes qualitatively at the first few virtual
cutoffs.  Cladding modes with $\Ulm< \tilde{U}_{l,1'}$ have essentially 
no intensity within the core.
(As with  the classic two layer model, $\tilde{U}_{1,1'}
\equiv 0$. Therefore, $l=1$ modes always guide some light within the core region.)
When the normalized wave-vector satisfies $\tilde{U}_{l,1'}<U<\tilde{U}_{l,2'}$, 
HE modes ($P_{lm}\approx-1$) have one maximum within the core.  In
contrast, EH modes ($P_{lm}\approx 1$) have a minimum at the center and
have significant field contributions only near the core-cladding boundary. 
Thus the spatial distribution of energy is distinct for HE versus EH.
The polarization of each mode in this regime is quite complex and far from uniform across the fiber.

The propagation of modes with $U>\tilde{U}_{l,2}$ is more determined by total
internal reflection at the cladding/surrounding interface than the core/cladding
interface. 
In this region we have either $|\Plm| \approx 0$ (HE), 
or $|\Plm| \gg 1$ (EH), which is indicative of modes
for which  the energy in the longitudinal components resides almost purely in $\Hz$ (HE)
or purely in $\Ez$ (EH).  As one or other longitudinal component is close to zero, 
consequently the transverse field of the HE modes eventually become close to 
purely azimuthally polarized (quasi-TE) and that of the EH modes becomes
strongly radially polarized (quasi-TM)~\cite{steel:2004p8122}.

\section{Calculation of the spectra}\label{sec:cmt}
We now turn to the detailed explanation of the form of the cladding mode spectra,
calculating the resonant wavelengths, the coupling coefficients
and finally the complete transmission spectra.

\subsection{Resonance wavelengths}
One of the challenges of computing FBG spectra of strong cladding mode coupling is
that it usually occurs over large bandwidth ranging from 20 to several hundred
nanometers.  
All resonances that result from the reflection of the incoming mode \mHE{1}{1} into
cladding modes $(l,m)$ by  the grating of period $\Lambda$ 
are defined by the phase matching condition 
\begin{equation}\label{eq:phasematching}
	\beta_{11} \pm \beta_{lm} - 2\pi/\Lambda=0,
\end{equation}
Taking all the propagation constants as positive, the upper sign holds  for FBGs, 
and the lower sign for long period gratings.

Since Eq.~\eqref{eq:phasematching} involves the wavelength implicitly through the effective
indices of the modes, computing the resonance wavelengths requires solving 
the dispersion relation for hundreds of modes at finely discretized wavelengths.
Within a bandwidth of 30~nm, the material dispersion of the fiber can usually be
neglected as it is of the order of $10^{-4}$  and affects both core and
cladding refractive indices in a similar fashion. Therefore the index contrast $\ncore-\nclad$ varies only weakly.
However, the waveguide dispersion is significant and must be accounted for.  
Solving Eq.~\eqref{eq:phasematching} exactly is relatively straightforward, but considerable care is required in solving
the dispersion relation~\eqref{eq:disprel} to find all the required effective indices $\neff_{lm}$ at all wavelengths, 
since some of the EH modes appear as resonances as narrow as one part in a million.

If the intention is to identify particular modes in a measured spectrum 
rather than precisely predict their location in advance,
a more convenient approximate approach that provides additional insight is to solve the dispersion relation
at the Bragg wavelength $\lambda_0$
and  extrapolate the waveguide dispersion of the propagation
constants using the relation~\cite{Saleh:1991p1302}
\begin{equation}  \label{eq:lpmodes}
	\beta_{lm}(\lambda)\approx \left( \nclad^2 \left(\frac{2\pi}{\lambda}\right)^2 - A_{lm} \right)^{1/2} \textnormal{with} \qquad
	A_{lm}=\frac{(l+2m+1/2)^2\pi^2}{4a_{2}^2},
\end{equation}
where $A_{lm}$ is a mode-dependent dispersion-free constant. 
This relation applies for the two-layer model of a fiber in air ($n_3=1$)
in the LP approximation and the regime of high $V$ number (for the cladding modes $V\approx 270$).
These LP resonances represent the HE/EH doublets for resonances of high radial number $m$ described earlier.  
Although this approximation neglects the vectorial mode splitting and is rather inaccurate for low $m$, it is helpful
for understanding the degeneracies with respect to the azimuthal number $l$: 
because of the $l+2m$ term, two sets of
propagation constants exist, one for odd $l$ and one for even $l$ halfway between.  

\begin{figure}[htbp]
	\centerline{ \includegraphics[width=\textwidth]{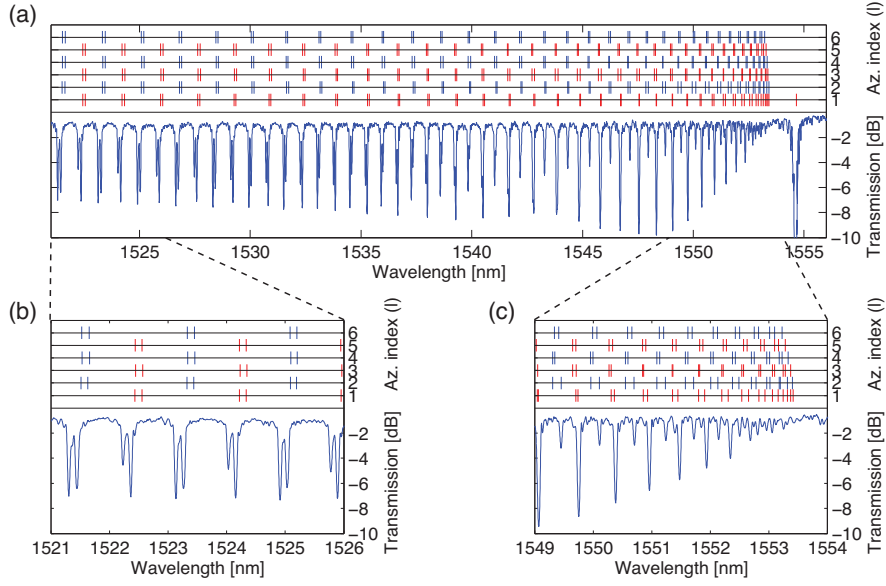}}
	\caption{Measured cladding mode spectra in transmission (bottom) and computed HE and EH resonances (top) 
for odd (red) and even (blue) azimuthal order $l$. 
(a) Complete measured wavelength range; (b) low-and (c) high-wavelength range in more detail.}
	\label{fig:labelled_spectra}
\end{figure}

In order to precisely predict the resonance wavelengths using Eq.~\eqref{eq:lpmodes},
for each mode the dispersion free constant 
\begin{equation}
\tilde{A}_{lm}=(n_2^2-\neff_{lm}^2)\left(\frac{2\pi}{\lambda_0}\right)^2,
\end{equation} 
was computed by solving the  exact three-layer model
for $\neff_{lm}(\lambda_0)$ at just the Bragg wavelength $\lambda_0$.
Substituting  $\tilde{A}_{lm}$ for $A_{lm}$ in~\eqref{eq:lpmodes} 
and combining that equation with~\eqref{eq:phasematching},
we obtain a quadratic equation for the wavelength of the $(l,m)$ resonance
\begin{equation}\label{eq:reswavelength}
\lambda_{lm}=\frac{ \frac{\neff_{11}}{\Lambda}\pm \left[\left(\frac{\neff_{11}}{\Lambda}\right)^2
    +\frac{1}{4\pi^2}\left(\tilde{A}_{lm}+\left(\frac{2\pi}{\Lambda}\right)^2\right)\left(\neff_{lm}^2-(\neff_{11})^2\right)\right]^{1/2}}
{\frac{1}{4\pi^2}\left(\tilde{A}_{lm}+\left(\frac{2\pi}{\Lambda}\right)^2\right)}.
\end{equation}
The positive root gives the wavelength for coupling into counter-propagating
modes.  The negative root yields the resonances for coupling into
co-propagating modes and is therefore neglected in the following treatment.
We have checked that this extrapolation procedure provides results in very 
close agreement with the exact approach.
Figure~\ref{fig:labelled_spectra} depicts a similar transmission spectrum
to that in Fig.~\ref{fig:firstpeaks}, together with the three-layer multimode resonances, 
computed according to the steps just described.  
The modes were calculated at $\lambda_0=1555$ nm. As expected from the LP
approximation in Eq.~\eqref{eq:lpmodes}, the
numerical solution also yields two sets of nearly-degenerate resonance
doublets, one for odd $l$ (red lines) and one for even $l$ (blue lines). 
For given $l$, each doublet  corresponds to an HE/EH pair.
The alignment of the measured and calculated resonances is good.
For low wavelengths (Fig.~\ref{fig:labelled_spectra}(b)), there is a small wavelength shift
compared to the measured spectra which arises because of the neglected material dispersion,
however the splitting of the HE/EH doublets is in very close agreement. 
In this range, the small difference in the propagation constant
$\beta_{HE}-\beta_{EH}$ of the HE/EH doublets is constant, as expected from
the standard 2-layer hybrid mode solution applied to the cladding-air boundary \cite{Snyder:1978p2682}. 
This is not the case for cladding resonances closer to the Bragg peak
(Fig.~\ref{fig:labelled_spectra}(c)): $\beta_{HE}-\beta_{EH}$ becomes
very small in the vicinity of a virtual cut-off.  
It can also be seen that in this regime, the doublets of a different azimuthal order are shifted such that the
EH resonance of a lower $l$ is in most but not all cases degenerate to the HE resonance for $l+2$. The use of LP modes $\mbox{\mLP{\ell}{m}}= \mbox{\mHE{\ell+1}{m}}+
\mbox{\mEH{\ell-1}{m}} $ still leads
to surprisingly good results for coupling strength of lower $m$ cladding modes
\cite{Hewlett:1996p2659}, since the contribution of higher $l$ modes cancels out in transversally homogeneous FBGs as shown in the following sections.

\subsection{Computation of coupling constants and spectral response}\label{sec:cme}
The interaction of the modes with the FBG can be
computed by solving the coupled mode equations as established by
Kogelnik~\cite{Kogelnik:1979p3485}. The unperturbed mode fields of the fiber and their 
propagation constants have been calculated beforehand as described in the preceding sections.
Here, we use the coupled mode equations
adapted for FBGs by Erdogan~\cite{Erdogan:1997p833}. This set of differential
equations is determined by the coupling constants, which are obtained by
evaluating an overlap integral of the interacting modes with the perturbation.
In the case of FBGs, the transverse coupling constant that governs the coupling
of the incoming
fundamental mode \mHE{1}{1} to any other reflected mode $(l, m)$ is computed with
\begin{equation}\label{eq:coupling_integral}
\kappa_{lm}=\frac{4}{\pi \nu}\cdot \frac{\omega }{4} \cdot 
    \int_0^{2\pi} \textnormal{d}\phi \int^{\infty}_0  \textnormal{d}r\, r  \Delta{\epsilon}(r,\phi)\, \textbf{E}_{11}^\textnormal{T}\cdot\textbf{E}_{lm}^\textnormal{T*},
\end{equation}
where only the transverse electric fields $\textbf{E}^\textnormal{T}_{lm}$ are
evaluated.  (The coupling associated with the longitudinal field components
is orders of magnitude smaller and can be neglected.)
The perturbation of the dielectric constant $\Delta \epsilon
(r,\phi)$ is induced here by refractive index modifications of the grating. The
grating is modeled as periodic but non-sinusoidal in the propagation
direction. For simplicity, a rectangular profile with a 50\% duty cycle is assumed, 
and the factor $4/(\pi \nu)$ accounts for the Fourier component at 
reflection order $\nu$. The investigated FBG reflects in second order ($\nu=2$).
To alleviate the need to compute a system of several hundred coupled mode
equations (one for each cladding mode), 
the synchronous approximation is used in which  all modes not near resonance are neglected.  
Solving the remaining equations yields wavelength dependent amplitudes, from which
reflection and transmission spectra can be accurately computed
~\cite{Erdogan:1997p833,Erdogan:1996p855}.  

\subsection{Fundamental coupling properties of the modes}\label{sec:fun}
Before characterizing the modification term $\Delta \epsilon$ in Eq.~\eqref{eq:coupling_integral}, it is
useful to elaborate on the kernel 
$K=\Delta \epsilon(r,\phi)  \textbf{E}_{11}^\textnormal{T}\cdot\textbf{E}_{lm}^\textnormal{T*}$ of the 
coupling constant integral, since this can be evaluated before integrating with the help of the results of section \ref{sec:classify}.  
Here, the incoming mode is always the \mHE{1}{1} core mode with
$\varphi=0$.  This mode couples with any other other hybrid mode,
characterized by $P_{lm}$ and $\ulm=(2\pi/\lambda)\sqrt{\ncore^2-\neff^2}$ with 
\begin{multline} \label{masterkernel}
K(\varphi) = \Delta \epsilon(r,\phi) \,  E_{11} E_{lm} \frac{\uii \ulm}{2} J_0(\uii r) \left\{
 (1-P_{lm})\Jlmlm \cos[(l-1) \phi +\varphi ] \right. \\
 \left. -(1+P_{lm}) \Jlplm \cos[(l+1) \phi + \varphi] \right\}
\end{multline}
as the kernel.  Note here, that the core mode can couple to modes of both sets of polarization
orientation, determined by $\varphi$.  We refer to the parallel kernel $K^\parallel=K(\varphi=0) $ 
for coupling to modes with the same azimuthal dependence as the launched core mode,
and the perpendicular kernel $K^\perp=K(\varphi=-\pi/2)$ for coupling to the orthogonal set of modes.

If the FBG extends homogeneously over the whole cross section of the fiber core and its
surrounding, Eq.~\eqref{eq:coupling_integral} would be equivalent to the
orthogonality relation of the modes, evaluating to zero for any mode $(l,m)$
other than the core mode \mHE{1}{1}.  In general, coupling between modes of
different indices occurs if these conditions are violated.  In conventional FBGs 
for example, the grating is only formed in the photosensitive
core and not the cladding. 
This case corresponds to a trivial integration over $\phi$ and any terms with a
sine or cosine cancel out.  Thus, only the first term of
Eq.~\eqref{masterkernel} remains for $l=1$ and arbitrary $m$, while the
perpendicular kernel $K^\perp$ always sums to zero.  This case is described
in detail in~\cite{Erdogan:1997p833}.  In particular, coupling to higher order
modes with $l>1$ is only possible if the refractive index of the FBG is
inhomogeneous or asymmetric. The strength of such coupling is influenced by the
cross section of the FBG through its non-uniformity and localization.

The classification of the fiber modes as discussed in section \ref{sec:classification} is reflected in the kernel, too. 
For the cladding modes below the $\tilde{U}_{l,2'}$ cutoff,
the strength of the coupling coefficient has a pronounced alternating character
as a function of $m$ due to the hybrid mode parameter $P_{lm}\approx \pm 1$: The $(1-P_{1m})$ term in 
Eq.~\eqref{masterkernel} is responsible for the strong selective character.
While for HE modes the $\Jlmlm$ term dominates, it is the $\Jlplm$ term for
EH.  In contrast, cladding modes beyond the $\tilde{U}_{l,2'}$ cut-off have 
quasi-TM/TE core fields.  Because of this, $P_{lm}\neq \pm 1$ and HE and EH modes differ less in 
their coupling behavior.

\subsection{Coupling coefficients and spectra}
We now proceed to calculate the coupling coefficients and complete transmission spectra
for highly-localized FBGs.  In our experiments, the FBG consists of femtosecond pulse induced micro voids (see Fig.~\ref{fig:mod_sketch} and Ref.~\cite{Jovanovic:2009p2656},) 
that are approximately prolate spheroids (with semi-major axis 0.95~\micron and
semi-minor axis 0.2~\micron,). In the following, the densified shell around the microvoid is
neglected, since its influence on the overall coupling is small compared to the
void itself \cite{Jovanovic:2009p2656}.  
To reduce the geometry to the form of Eq.~\eqref{eq:coupling_integral},
we treat the grating as a periodic square modulation with an elliptical cross-section and an index
contrast obtained by averaging the local index of the true structure 
over a whole period.  
Assuming this, 
the perturbation of the dielectric constant is 
\begin{equation}
	\Delta \epsilon (x,y)= \epsilon_0 (n_{mod}^2-n_1^2) \theta (x,y)
\end{equation}
where $\theta (x,y)=1$ inside the ellipse and $0$ elsewhere, and $n_{mod}=1.325$~\cite{Jovanovic:2009p2656}. 

\begin{figure}[t]
\begin{center}
\includegraphics[width=.8\textwidth]{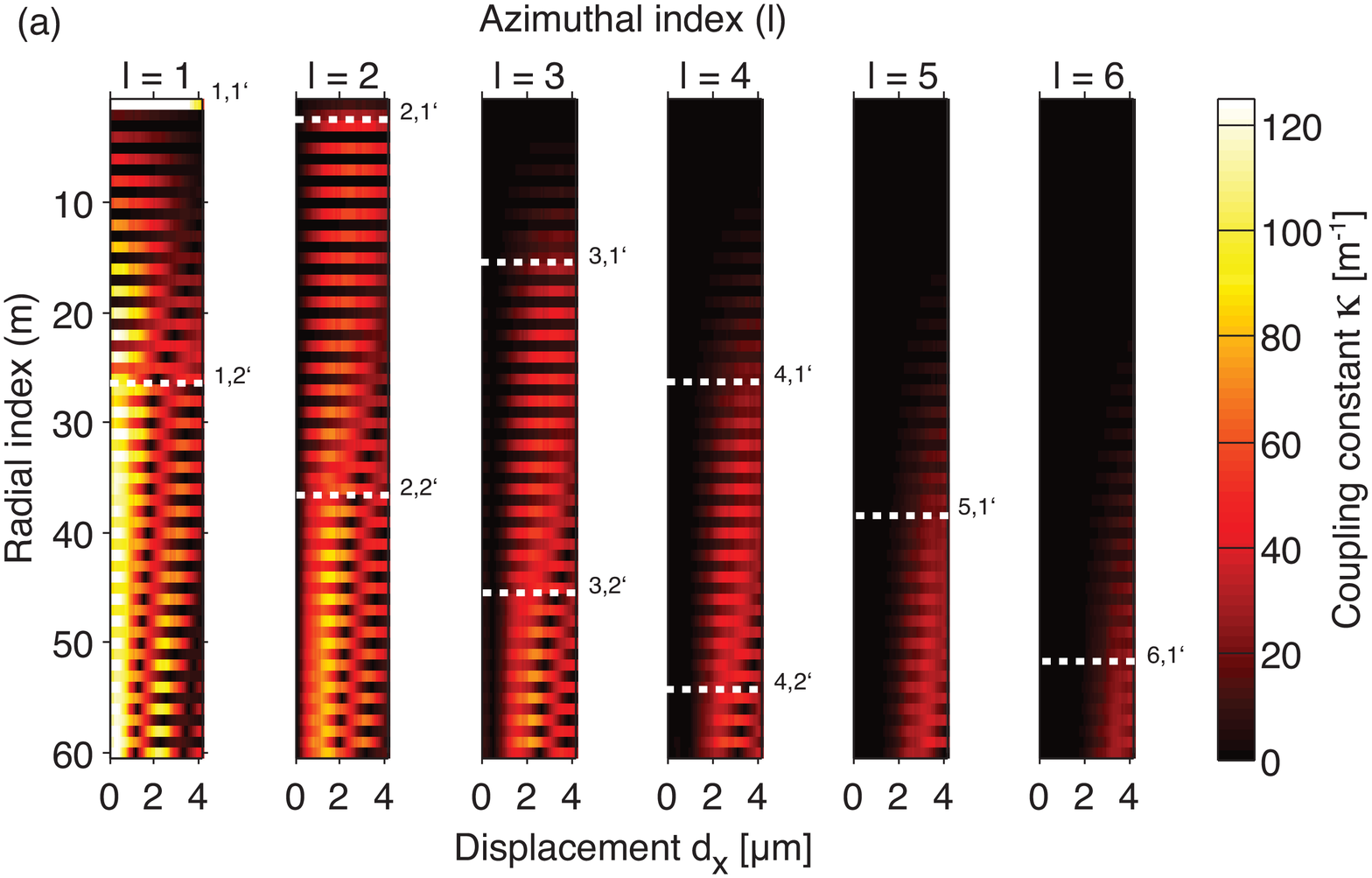} \\
\includegraphics[width=.8\textwidth]{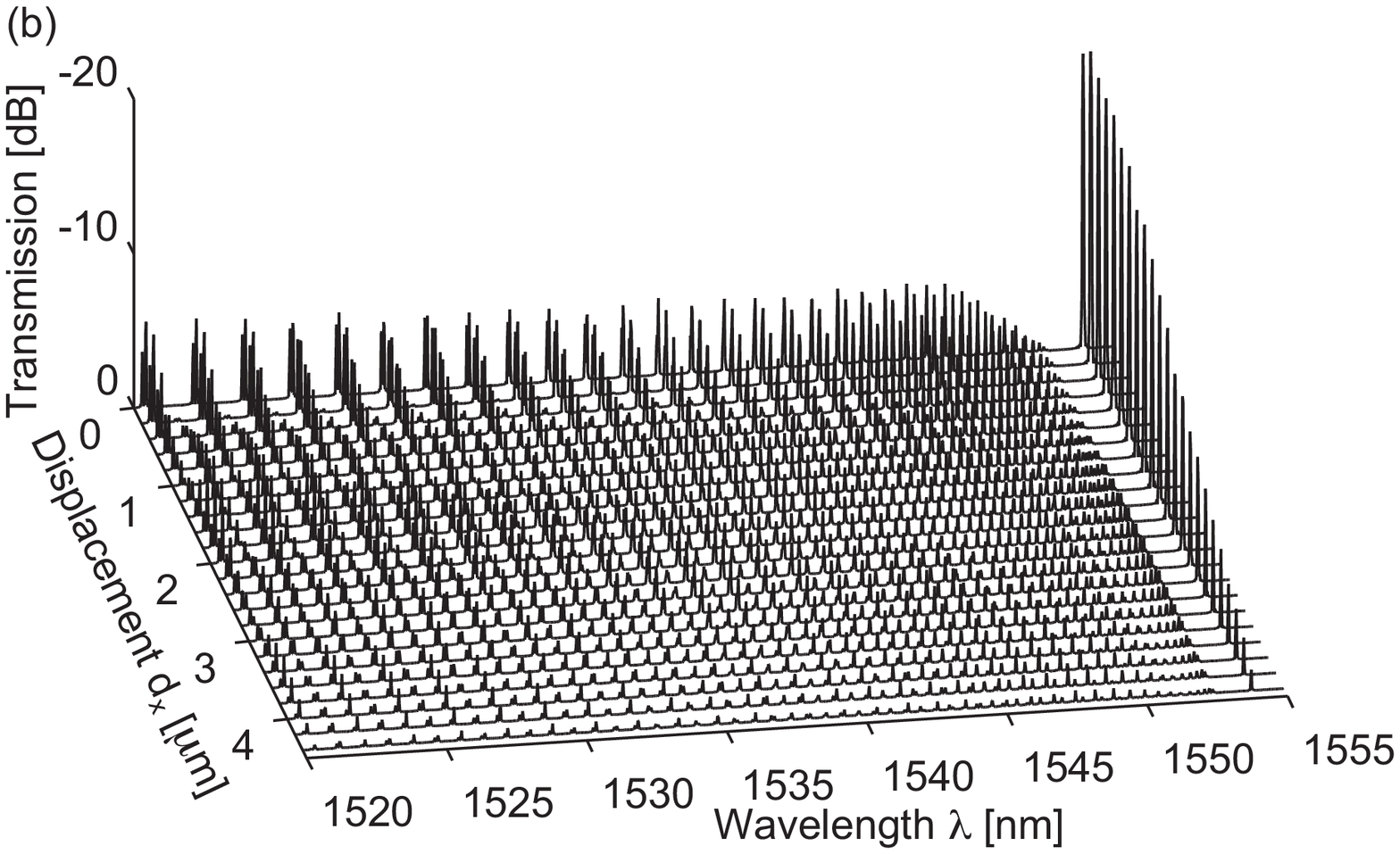}
\caption{(a) 
Core-cladding mode coupling coefficients for different displacements
$d_x$ of the microvoid in $x$-direction from the center of the fiber core.
Coupling constants for $l=1$ to $l=6$ are plotted, $m$ is ascending from top
to bottom.  Dotted lines indicate the virtual cutoffs.
(b) Computed transmission spectra for different displacement of the modification in the $x$-direction.}
\label{fig:coupling_vs_decenter_x}
\end{center}
\end{figure}


In our fabricated gratings we can position the voids arbitrarily in the core (although there is some variation in the void displacement along the core due to vibration in the ferrule during
the writing process).  The strength of the cladding mode coupling strongly depends on small
displacements of the micro voids from the central axis of the fiber (results presented below
in Fig.~\ref{fig:spectra_comparison}). For inhomogeneous FBGs, 
the launch polarization has a significant influence on the overall coupling strength
because of coupling to both sets of modes (Eq.~\eqref{masterkernel}). These polarization dependent effects will
be discussed in more detail in a forthcoming publication. In the following calculations, we emphasize more the dependence of 
the coupling to the radial position of the highly localized FBG. We therefore choose the special case for which
 the launch polarization is aligned with the long axis of the modifications
and the modifications are only displaced along the $x$-axis.
The perpendicular kernels $K^\perp$ therefore cancel out and can be neglected.

The coupling integral was evaluated for displacements $d_x$ in the
$x$-direction, ranging from $0$ to the core cladding boundary at $4.15$~\micron. The resulting coupling constants for the parallel kernel $K^\parallel$ are plotted in Fig.~\ref{fig:coupling_vs_decenter_x}(a) as a function of radial mode
order $m$ and displacement $d_x$ for azimuthal orders $l=$1--6.  The horizontal
white dotted lines show the locations of the virtual cutoffs for
each value of $l$.  Since resonances above $\tilde{U}_{7,1'}$ lie beyond the
wavelength range of our measured spectra, we expect negligible coupling to modes of $l>6$, so that the range $l=$1--6 is appropriate. 
All of the concepts of the preceding section are present in these plots: 
there is almost no coupling below the $\tilde{U}_{l,1'}$
cutoff, pure HE coupling for modes below $\tilde{U}_{l,2'}$ (as indicated by the bright red bands
for odd $m$ (HE) and dark bands for the even $m$ (EH) modes); and reduced contrast between
the HE and EH selective coupling above $\tilde{U}_{l,2'}$, when the nature of the modes changes to 
quasi-TE and quasi-TM. In agreement with~\cite{Erdogan:1997p833} and the earlier symmetry
argument, there is no coupling into modes of $l\neq1$ if the modification is
centered ($d_x=0$). 
Figure~\ref{fig:coupling_vs_decenter_x}(b) displays the corresponding spectra
calculated from the coupling constants of \ref{fig:coupling_vs_decenter_x}(a) as outlined in section \ref{sec:cme}. 
The even $l$ resonances appear in the spectra in between the odd $l$ resonances (see Eq.~\ref{eq:lpmodes}).  
Most striking is the sensitivity of coupling to $l=2$ modes: even a small displacement $d_x\approx$~0.4~\micron
of the modification suffices for measurable coupling strength. 
The transmission spectra do not allow
us to distinguish the strength of coupling to modes with $l$ of the same parity,
but the maps of coupling coefficient in (Fig~\ref{fig:coupling_vs_decenter_x} a)) indicate that measurable coupling to $l=3$ and $l=4$ modes can only be observed if the modification 
is at least 1~\micron from the center. 

\begin{figure}[htbp]
\begin{center}
\includegraphics[width=1\textwidth]{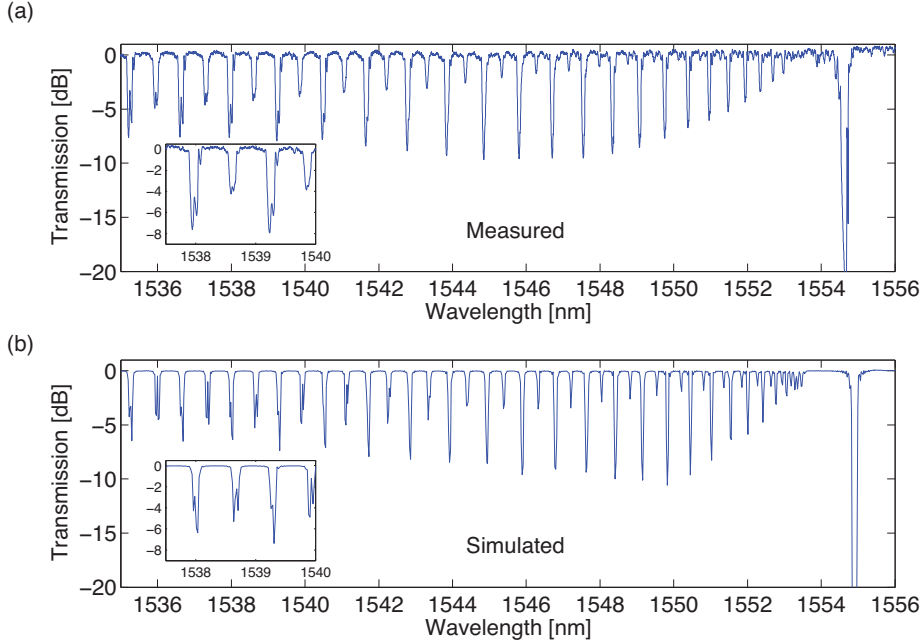}
\caption{Measured (a) and computed (b) transmission spectra for the micro void at $d_x=0.8$ $d_y=0.7$. The insets show the EH/HE doublets.}
\label{fig:spectra_comparison}
\end{center}
\end{figure}


\section{Discussion}
In Figure~\ref{fig:spectra_comparison} we compare measured and calculated transmission spectra for one FBG with the light launched perpendicular polarized to the fiber axis. 
The positions of the FBGs within the fiber core were estimated from microscope 
images.  Due to fiber vibration in the ferrule, the micro voids are not uniformly distributed along the FBG, 
and the displacement values used in calculations are the maximum measured
values.  In the first case (Fig.~\ref{fig:spectra_comparison}(a) and~(b), the
spectrum was simulated for $d_x=0.8$~\micron and $d_y=0.7$~\micron.
Both in experiment and simulation, the polarization
direction was along the long axis of the micro void, parallel to the $y$-
axis. Because of the displacement in $y$-direction, the perpendicular coupling constants
($\varphi=-\pi/2$ in Eq.~\ref{masterkernel}) had to be considered as well.
The computed spectra closely agree with the experimentally obtained spectra.
The slight difference in the envelope can be attributed to the restriction of
the model to account only for uniform displacement.  
(Note that we have also performed coupling calculations using a core mode that was calculated using the finite element
method to include the mean index perturbation associated with the grating modifications.  This
makes a negligible change to the transmission spectra.)

Due to the displacement from the center and the therefore
asymmetric cross-section of the FBG, coupling to higher order modes occurs. This is clearly visible in the
second comb of resonances in between the $l=1$ resonances, which corresponds to
the even $l$ modes [see Eq.~\eqref{eq:lpmodes}]. Additionally, with the aid of imaging the reflected cladding
modes, we were able to determine that interference
between degenerate modes of different azimuthal order took place and therefore
that higher order azimuthal modes were significantly excited. This will be the subject of a future publication.

Consistent with the results in Fig.~\ref{fig:labelled_spectra}, the low order resonances in
Fig.~\ref{fig:spectra_comparison}(a) appear as single
peaks but doublets appear at higher $m$ (shorter wavelengths). 
 The single peaks correspond to pure HE modes.  The EH modes are formally present in these peaks, but as seen
in Fig.~\ref{fig:coupling_vs_decenter_x}(a), the coupling is too weak for them
to be observed. While the single peaks show almost no polarization dependence, the strength of the doublet peaks heavily depends on how the axis of polarization is orientated with respect to FBG. The experimental and theoretical results concerning this aspect will also be presented in a future publication.


\section{Conclusion}
In this paper we investigated the core-cladding coupling behavior of FBGs that
have transversely inhomogeneous refractive index profiles.  For that purpose,
we wrote intra core FBGs in step index fibers based on the point-by-point
approach.  The micro voids formed within the index modifications that the
grating comprises, are much smaller than the core and therefore allow for an
accurate excitation of particular modes.  Here, even $l$
resonances served as a strong test for the centeredness of a FBG. 
Our findings were validated by a rigorous calculation of the coupling constants.  Accurate
spectra of transversely inhomogeneous FBGs could be reproduced, which
predicted many of the transmission characteristics of such FBGs, that have only
been described qualitatively before. The necessary theoretical foundation for a detailed investigation
of polarization dependent features could be established. 

We have demonstrated efficient mode converters to odd and even $l$ modes, realized with off-centered, 
highly localized FBGs. Such intra-core gratings allow for specifically targeting coupling to cladding modes
of higher azimuthal index $l$.  We believe these findings will enable
many new fiber sensor designs, that rely on asymmetrically structured fiber
cores. The efficient core-cladding mode conversion presented here can also
 be exploited for in-fiber elements that allow for the
tailored excitation of higher-order mode fibers \cite{Ramachandran:2008p8304}.
Our extensive treatment can be easily transferred to fibers that carry
multiple modes in the core for design and calculation of core-mode
converting gratings. 

\section{Acknowledgments}
The financial support by the German Federal Ministry of Education and Research
(BMBF) under Contract No. 13N9687 is gratefully acknowledged. Jens Thomas also
acknowledges funding by the DAAD, grant D/0846673.
CUDOS is an Australian Research Council Centre of Excellence.

%

\appendix


\section{Dispersion relation}
According to \cite{Erdogan:1997p833,Tsao:1989p1975}, the dispersion for the three layer fiber with arbitrary integer azimuthal number $l$ is
\begin{subequations}
\label{eq:disprel}
\begin{equation}
\zeta_0=\zeta_0',
\end{equation}
where
\begin{equation}
\zeta_0=\frac{1}{\sigma}
\frac{\vlm \left(JK+\frac{\sigma^2  \vtwoone \vthreetwo }{\nclad^2 \acore \aclad}\right)p_l(\aclad)
-Kq_l(\aclad) + Jr_l(\aclad)-\frac{1}{\vlm }s_l(\aclad)}
{-\vlm \left(\frac{\vthreetwo }{\nclad^2\aclad}J-\frac{\vtwoone}{\ncore^2 \acore}K\right)p_l(\aclad)
+\frac{\vthreetwo}{\ncore^2\aclad}q_l(\aclad)+\frac{\vtwoone}{\ncore^2 \acore}r_l(\aclad)}
\end{equation}
\begin{equation}
\zeta_0'=\sigma \frac{ \vlm  \left( \frac{\vthreetwo}{\aclad}J-\frac{\nair^2 \vtwoone}{\nclad^2\acore}K \right)p_l(\aclad)
-\frac{\vthreetwo}{\aclad}q_l(\aclad)-\frac{\vtwoone}{\acore}r_l(\aclad) }
{\vlm \left(\frac{\nair^2}{\nclad^2}JK+\frac{\sigma^2 \vtwoone\vthreetwo}{\ncore^2\acore}\right)p_l(\aclad)
-\frac{\nair^2}{\ncore^2}Kq_l(\aclad)+Jr_l(\aclad)-\frac{\nclad^2}{\ncore^2\vlm }s_l(\aclad)}.
\end{equation}
\end{subequations}
Here, we have defined
\begin{align}
\ulm ^2 =(2\pi/\lambda)^2(\ncore ^2-\neff^2) , \qquad
\vlm ^2 =(2\pi/\lambda)^2(\nclad ^2-\neff^2) , \qquad
\wlm ^2=(2\pi/\lambda)^2(\neff^2-\nair ^2),
\end{align}
and
\begin{align}
\sigma&=il\neff , \qquad \vtwoone =\frac{1}{\vlm ^2}-\frac{1}{\ulm^2}, 
  \qquad \vthreetwo=\frac{1}{\wlm ^2}+\frac{1}{\vlm ^2}, \qquad
J=\frac{J_l'(\ulm \acore)}{\ulm J_l(\ulm \acore)} , \qquad
K=\frac{K_l'(\wlm \aclad)}{\wlm K_l(\wlm \aclad)}.
\end{align}
The dispersion relation includes Bessel-functions of the first and second kind $J_n$ and $N_n$ as well as modified Bessel-functions $K_n$ of the second kind. In addition, the products 
\begin{subequations}
\begin{align}
p_l(r)&=J_l(\vlm r)N_l(\vlm \acore)-J_l(\vlm \acore)N_l(\vlm r),\\
q_l(r)&=J_l(\vlm r)N_l'(\vlm \acore)-J_l'(\vlm \acore)N_l(\vlm r),\\
r_l(r)&=J_l'(\vlm r)N_l(\vlm \acore)-J_l(\vlm \acore)N_l'(\vlm r),\\
s_l(r)&=J_l'(\vlm r)N_l'(\vlm \acore)-J_l'(\vlm \acore)N_l'(\vlm r),
\end{align}
\end{subequations}
are used, where the prime stands for differentiation with respect to the total argument. 

\section{Fields}
Here we provide the expressions for the fields in the cladding and air regions,
Inside the cladding ($\acore \leq r \leq \aclad$):
\begin{subequations}
\begin{alignat}{2}
\Ez^\cl   & =&   & \Ecllm \Ccllm   \frac{\vlm ^2\sigma}{\nclad^2\beta l}\left[G_2p_l(r)-\frac{\nclad ^2\zeta_0}{\ncore^2\vlm }q_l(r)\right]   \sinlp \, \eibzt \\
\Er^\cl   & =& i & \Ecllm \Ccllm   \left( -\frac{lF_2}{r}p_l(r) + \frac{l}{\vlm r}q_l(r)-\frac{\sigma}{l \nclad^2}\left[\vlm G_2 r_l(r)-\frac{\nclad^2\zeta_0}{\ncore^2}s_l(r)\right]\right) \sinlp \, \eibzt \\
\Ephi^\cl & =&-i & \Ecllm \Ccllm   \left( \frac{\sigma}{\nclad^2}\left[\frac{G_2}{r}p_l(r)-\frac{\nclad^2\zeta_0}{\ncore^2\vlm r}q_l(r)\right] +\vlm F_2r_l(r)-s_l(r)\right) \coslp \, \eibzt  
\end{alignat}
\end{subequations}
\begin{subequations}
\begin{alignat}{2}
\Hz^\cl   & =& i  & \Ecllm \Ccllm \frac{1}{Z_0}  \frac{\vlm ^2\sigma}{\beta l}\left[F_2p_l(r)-\frac{1}{\vlm }q_l(r)\right] \coslp \, \eibzt  \\	
\Hr^\cl   &= & -  & \Ecllm \Ccllm  \frac{1}{Z_0}\left(l \frac{G_2}{r}p_l(r) - l \frac{\nclad^2\zeta_0}{\ncore^2\vlm r}q_l(r) - \frac{\sigma}{l}\left[\vlm F_2r_l(r)-s_l(r)\right]\right) \coslp \, \eibzt  \\
\Hphi^\cl & =& -  & \Ecllm \Ccllm  \frac{1}{Z_0}\left(\sigma\left[\frac{F_2}{r}p_l(r)-\frac{1}{\vlm r}q_l(r)\right]-\vlm G_2r_l(r)+\frac{\nclad^2\zeta_0}{\ncore^2}s_l(r)\right) \sinlp \, \eibzt 
\end{alignat}
\end{subequations}
with the terms
\begin{equation}
C_{lm}=\frac{\pi \acore \ulm ^2 J_l(\ulm  \acore)}{2}, \qquad
F_2=J-\frac{u_{21}\sigma\zeta_0}{\ncore^2\acore}, \qquad G_2=\zeta_0J+\frac{\vtwoone \sigma}{\acore}.
\end{equation}
In the air region, ($r > \aclad$) the fields are
\begin{subequations}
\begin{alignat}{2}
\Ez^\cl   & =&-  & \Ecllm \Ccllm  \frac{\vlm ^2}{K_l(\wlm \aclad ) \wlm }  \frac{\sigma G_3 \wlm }{\beta l \nair^2} K_l(\wlm r) \sinlp \, \eibzt \\
\Er^\cl   & =&- i & \Ecllm \Ccllm  \frac{\vlm ^2}{K_l(\wlm \aclad ) \wlm}                \left[l \frac{F_3}{\wlm r} K_l(\wlm r) + \frac{G_3 \sigma}{l \nair^2}  K_l'(\wlm r) \right]   \sinlp \, \eibzt \\
\Ephi^\cl & =&- i & \Ecllm \Ccllm  \frac{\vlm ^2}{K_l(\wlm \aclad ) \wlm}                \left[ F_3 K_l'(\wlm r) + \frac{G_3 \sigma}{\wlm r \nair^2  }  K_l(\wlm r) \right]   \coslp \, \eibzt 
\end{alignat}
\end{subequations}
\begin{subequations}
\begin{alignat}{2}
\Hz^\cl   & =&-i & \Ecllm \Ccllm  \frac{1}{Z_0}\frac{\vlm ^2}{K_l(\wlm \aclad ) \wlm  } \frac{\sigma F_3 \wlm}{\beta l}  \coslp \, \eibzt \\
\Hr^\cl   &= &-   & \Ecllm \Ccllm  \frac{1}{Z_0}\frac{\vlm ^2}{K_l(\wlm \aclad ) \wlm }               \left[\frac{l G_3}{r \wlm} K_l(\wlm r) -\frac{F_3 \sigma}{l} K_l'(\wlm r) \right] \coslp \, \eibzt \\
\Hphi^\cl & =&-   & \Ecllm \Ccllm  \frac{1}{Z_0}\frac{\vlm ^2}{K_l(\wlm \aclad ) \wlm}                \left[- G_3 K_l'(\wlm r) +\frac{F_3 \sigma}{\wlm r} K_l(\wlm r) \right] \sinlp \, \eibzt 
\end{alignat}
\end{subequations}
with
\begin{align}
F_3=-F_2 p_l(\aclad) + \frac{1}{\vlm}q_l(\aclad) , \qquad
G_3=-\frac{\nair^2}{\nclad^2} \left[ G_2 p_l(\aclad) - \frac{\nclad^2 \zeta_0}{\ncore^2 \vlm} q_l(\aclad) \right].
\end{align}


\begin{thebibliography}{10}
\newcommand{\enquote}[1]{``#1''}

\bibitem{Erdogan:1997p833}
T.~Erdogan, \enquote{Cladding-mode resonances in short- and long-period fiber
  grating filters,} J. Opt. Soc. Am. A \textbf{14}, 1760--1773 (1997).

\bibitem{Mizrahi:1993p798}
V.~Mizrahi and J.~E. Sipe, \enquote{Optical properties of photosensitive fiber
  phase gratings,} J. Lightwave Technol. \textbf{11}, 1513--1517 (1993).

\bibitem{Jauregui:2004p3427}
C.~J{\'a}uregui, A.~Quintela, and J.~M. L{\'o}pez-Higuera,
  \enquote{Interrogation unit for fiber bragg grating sensors that uses a
  slanted fiber grating,} Opt. Lett. \textbf{29}, 676--678 (2004).

\bibitem{Erdogan:1996p855}
T.~Erdogan and J.~E. Sipe, \enquote{Tilted fiber phase gratings,} J. Opt. Soc.
  Am. A \textbf{13}, 296--313 (1996).

\bibitem{Chan:2007p7575}
C.-F. Chan, C.~Chen, A.~Jafari, A.~Laronche, D.~J. Thomson, and J.~Albert,
  \enquote{Optical fiber refractometer using narrowband cladding-mode resonance
  shifts,} Appl. Opt. \textbf{46}, 1142--1149 (2007).

\bibitem{Zhou:2006p3488}
K.~Zhou, L.~Zhang, X.~Chen, and I.~Bennion, \enquote{Optic sensors of high
  refractive-index responsivity and low thermal cross sensitivity that use
  fiber bragg gratings of 􏰁80$\,^{\circ}$ tilted structures,} Opt. Lett.
  \textbf{31}, 1193--1195 (2006).

\bibitem{Mizunami:2000p2676}
T.~Mizunami, T.~Djambova, T.~Niiho, and S.~Gupta, \enquote{Bragg gratings in
  multimode and few-mode optical fibers,} J. Lightwave Technol. \textbf{18},
  230--235 (2000).

\bibitem{Guo:2009p7572}
T.~Guo, L.~Shao, H.-Y. Tam, P.~A. Krug, and J.~Albert, \enquote{Tilted fiber
  grating accelerometer incorporating an abrupt biconical taper for cladding to
  core recoupling,} Opt. Express \textbf{17}, 20651--20660 (2009).

\bibitem{Guo:2008p6002}
T.~Guo, A.~Ivanov, C.~Chen, and J.~Albert, \enquote{Temperature-independent
  tilted fiber grating vibration sensor based on cladding-core recoupling,}
  Opt. Lett. \textbf{33}, 1004--1006 (2008).

\bibitem{Martinez:2004p2878}
A.~Martinez, M.~Dubov, I.~Khrushchev, and I.~Bennion, \enquote{Direct writing
  of fibre bragg gratings by femtosecond laser,} Electron. Lett. \textbf{40},
  1170--1172 (2004).

\bibitem{Marshall:2010p8260}
G.~Marshall, R.~Williams, N.~Jovanovic, M.~J. Steel, and M.~J. Withford,
  \enquote{Point-by-point written fiber-bragg gratings and their application in
  complex grating designs,} Opt. Express \textbf{18}, 19844--19859 (2010).

\bibitem{Thomas:2007p80}
J.~Thomas, E.~Wikszak, T.~Clausnitzer, and U.~Fuchs, \enquote{Inscription of
  fiber bragg gratings with femtosecond pulses using a phase mask scanning
  technique,} Appl. Phys. A \textbf{86}, 153--157 (2007).

\bibitem{Jovanovic:2009p2656}
N.~Jovanovic, J.~Thomas, R.~J. Williams, M.~J. Steel, G.~D. Marshall,
  A.~Fuerbach, S.~Nolte, A.~T{\"u}nnermann, and M.~J. Withford,
  \enquote{Polarization-dependent effects in point-by-point fiber bragg
  gratings enable simple, linearly polarized fiber lasers,} Opt. Express
  \textbf{17}, 6082--6095 (2009).

\bibitem{Smelser:2005p41}
C.~W. Smelser, S.~J. Mihailov, and D.~Grobnic, \enquote{Formation of type i-ir
  and type ii-ir gratings with an ultrafast ir laser and a phase mask,} Opt.
  Express \textbf{13}, 5377--5386 (2005).

\bibitem{Snyder:1978p2682}
A.~Snyder and W.~Young, \enquote{Modes of optical waveguides,} J. Opt. Soc. Am.
  \textbf{68}, 297--309 (1978).

\bibitem{Hewlett:1995p2660}
S.~J. Hewlett, J.~D. Love, G.~Meltz, T.~J. Bailey, and W.~W. Morey,
  \enquote{Cladding-mode coupling characteristics of bragg gratings in
  depressed-cladding fibre,} Electron. Lett. \textbf{31}, 820--822 (1995).

\bibitem{Hewlett:1996p2659}
S.~Hewlett, J.~Love, G.~Meltz, T.~Bailey, and W.~Morey, \enquote{Coupling
  characteristics of photo-induced bragg gratings in depressed-and
  matched-cladding fibre,} Opt. and Quantum Electron. \textbf{28}, 1641--1654
  (1996).

\bibitem{Tsao:1989p1975}
C.~Tsao, D.~Payne, and W.~Gambling, \enquote{Modal characteristics of
  three-layered optical fiber waveguides: a modified approach,} J. Opt. Soc.
  Am. A \textbf{6}, 555--563 (1989).

\bibitem{steel:2004p8122}
M.~J. Steel, \enquote{Reflection symmetry and mode transversality in
  microstructured fibers,} Opt. Express \textbf{12}, 1497--1509 (2004).

\bibitem{Saleh:1991p1302}
B.~E.~A. Saleh, M.~C. Teich, and J.~W. Goodman, \enquote{Fundamentals of
  photonics,} Wiley pp. 272--309 (1991).

\bibitem{Kogelnik:1979p3485}
H.~Kogelnik, \enquote{Theory of dielectric waveguides,} Topics in Applied
  Physics \textbf{7}, 15--83 (1979).

\bibitem{Ramachandran:2008p8304}
S.~Ramachandran, J.~Fini, M.~Mermelstein, J.~W. Nicholson, S.~Ghalmi, and M.~F.
  Yan, \enquote{Ultra-large effective-area, higher-order mode fibers: A new
  strategy for high-power lasers,} Laser {\&} Photon. Rev. \textbf{2}, 429--447
  (2008).

\end{thebibliography}
\end{document}